# A simple KPFM-based approach for electrostatic-free topographic measurements: the case of MoS$_2$ on SiO$_2$

*Aloïs Arrighi[1,3], Nathan Ullberg[2], Vincent Derycke[2], Benjamin Grévin[1]*

[1,3]Univ. Grenoble Alpes, CNRS, CEA, IRIG-SyMMES, 38000 Grenoble, France

[2]Université Paris-Saclay, CEA, CNRS, NIMBE, LICSEN, 91191 Gif-sur-Yvette, France

Corresponding authors benjamin.grevin@cea.fr

[3]Institut Néel, CNRS, Univ. Grenoble-Alpes, 38042 Grenoble Cedex 09, France

# Abstract

A simple implementation of Kelvin probe force microscopy (KPFM) is reported that enables recording topographic images in the absence of any component of the electrostatic force (including the static term). Our approach is based on a close loop z-spectroscopy operated in data cube mode. Curves of the tip-sample distance as a function of time are recorded onto a 2D grid. A dedicated circuit holds the KPFM compensation bias and subsequently cut off the modulation voltage during well-defined time-windows within the spectroscopic acquisition. Topographic images are recalculated from the matrix of spectroscopic curves. This approach is applied to the case of transition metal dichalcogenides (TMD) monolayers grown by chemical vapour deposition on silicon oxide substrates. In addition, we check to what extent a proper stacking height estimation can also be performed by recording series of images for decreasing values of the bias modulation amplitude. The outputs of both approaches are shown to be fully consistent. The results exemplify how in the operating conditions of non-contact AFM under ultra-high vacuum, the stacking height values can dramatically be overestimated due to variations in the tip-surface capacitive gradient, even though the KPFM controller nullifies the potential difference. We show that the number of atomic layers of a TMD can be safely assessed, only if the KPFM measurement is performed with a modulated bias amplitude reduced at its strict minimum or, even better, without any modulated bias. Last, the spectroscopic data reveal that defects at the TMD/oxide interface can have a counterintuitive impact on the electrostatic landscape, resulting in an apparent decrease of the measured stacking height by conventional nc-AFM/KPFM compared to non-defective sample areas. Hence, electrostatic free z-imaging proves to be a promising tool to assess the existence of defects in atomically thin TMD layers grown on oxides.

# 1. Introduction

Since its introduction in the early 1990s[1], Kelvin Probe Force Microscopy (KPFM) has become one of the most widespread electrostatic variants of the Atomic Force Microscope (AFM). KPFM is based on the detection of the electrostatic force component due to a potential difference between the surface and the AFM cantilever's tip, and its subsequent minimization by a proper electric potential. Mapping this compensation bias yields access to the electrostatic landscape of a sample's surface. Depending on the type of sample under investigation, KPFM data can be analysed in terms of local work function variations, charge distributions, interface dipoles, or photo-induced charge populations.

Very early on, it became also evident that better height measurements would be performed by combining dynamic AFM modes with KPFM. In the case of heterogeneous materials, or samples nanostructured on purpose, local variations in the effective work function (or surface charge distributions) affect the electrostatic force (and its gradient), and hence the tip-sample distance regulation. In KPFM, these electrostatic-related topographic artefacts are mitigated by the active compensation of the tip-surface potential difference. This has been nicely illustrated in the pioneering work of Sadewasser and Lux-Steiner[2], who demonstrated that noncontact-AFM combined with KPFM allowed performing correct estimations of C60 sub-monolayers stacking height on highly oriented pyrolytic graphite. Almost twenty years after, there is no doubt about the relevance of this approach, which continues to contribute to the improvement of AFM-based techniques. For instance, electrostatic artefact compensation by KPFM has recently been applied to improve the performances of scattering scanning near-field optical microscopy[3].

It would however be a dangerous illusion to think that performing topographic measurements with an active KPFM loop provides an absolute guarantee against electrostatic-induced topographic artefacts. A well-known fact[4], unfortunately too often overlooked, is that the application of the modulated bias (used for the electrostatic force detection via a lock-in scheme) generates a static force component proportional to the tip-sample capacitive gradient. Consequently, topographic artefacts may exist if this last parameter displays spatial variations.

In some cases, it is reasonable to assume that this effect will not impact significantly the topographic profiles recorded by an nc-AFM. This should hold true, for instance, for the abovementioned C60/graphite benchmark: there is here no reason to expect significant variations of the tip-surface capacitive gradient, regardless of the material under the tip. By contrast, artefacts should be an unavoidable consequence of the capacitance variations for conducting layers deposited on insulating thin films. For instance, in the case of graphene flakes deposited on doped-silicon covered by silicon oxide (SiO2), the capacitance is either the one between the AFM tip and the backside doped silicon, or a serial concatenation of two capacitances[5] (between the tip and the flake, and between the flake and the doped silicon).

Transition metal dichalcogenides (TMDs) are another class of two-dimensional materials, which are very often processed on silicon oxide substrates as the active parts of opto-electronic devices such as field effect transistors and photodetectors[6]. Their optoelectronic properties are strongly dependent of the material thickness. In particular, many TMDs (such as $MoS_2$, $WS_2$, $MoSe_2$, $WSe_2$) exhibit an indirect to direct band gap transition when thinned down to the 2D monolayer. For a wide community of researchers, assessing correctly the stacking height (and consequently the number of atomic layers) of TMDs on oxide surfaces is an essential prerequisite to any basic or applied research.

Numerous studies use optical methods and spectroscopies[7] to determine the number of layers in TMD samples. The AFM has become a widespread basic characterization tool, and complementary investigations by AFM profiling are often performed. However, as noted by others[8], there exist a significant disparity between the results of published AFM studies on exfoliated TMD flakes on SiO2. The inadequacy of AFM to yield consistent stacking height values of 2D layers on oxide surfaces has also been noted in the case of graphene[9]. It has been rightly pointed out that electrostatic forces can misleadingly influence the stacking height measurement of 2D materials by AFM[8]. To address this problem, one may be tempted to apply an active KPFM loop to compensate the tip-surface potential difference. Unfortunately, due to capacitive-related artefacts, the cure could reveal itself worse than the disease. This is especially likely if one uses large modulated bias (Vac), since the static electrostatic force component scales with the product of the capacitance gradient by the square of the modulated bias amplitude[4].

In a recent work[10], Ritz and co-workers have developed a Kalman-filter based methodology that allows estimating the electrostatic influence due to the applied voltage modulation on the cantilever frequency shift. The imaging can then be performed on the basis of the sole topography-induced part of the frequency shift. Doing so, they nicely confirmed that the capacitive artefacts noticeably affect the AFM/KPFM topographic measurements performed on graphene flakes on silicon oxide

In this work, we follow an alternative strategy, which consists in tackling the root of the problem. The idea is to suppress completely the electrostatic force during the topographic measurement. This strategy is applied to the case of $MoS_2$ monolayers grown on silicon oxide by chemical vapour deposition. Prior to this, we show that the stacking height of a $MoS_2$ monolayer on $SiO_2$ can be accurately estimated by recording a series of AFM/KPFM images for decreasing values of the modulated bias. In a second step, we demonstrate how "electrostatic free" height measurements can be performed. Our approach is based on the implementation of a new distance-spectroscopy protocol operated in data cube mode. In each pixel, curves of the tip-sample distance (z) as a function of time are synchronously recorded with trigger events that i) drive a hold circuit which maintains the KPFM compensation potential to a constant value, and ii) drive a second circuit that cut off the modulated bias. Artefact-free topographic images are obtained by mapping the z-levels recorded when Vac is switched off. Moreover, by comparing the data acquired on two different samples, we show that defects at the TMD/oxide interface can counterintuitively impact the effective stacking height measured by conventional nc-AFM/KPFM. Beyond the possibility of performing correct topographic measurements, we thus show that our approach can be very useful to avoid misinterpretations, and assess more precisely the existence of defects at the TMD/oxide interface.

## 2. KPFM background

In KPFM, the tip and sample form a capacitive junction, and the attractive electrostatic force can be expressed as:

$$F_{el} = \frac{1}{2} C'_z (\Delta V)^2 \quad (1)$$

where C'z and ΔV are the tip-sample capacitive gradient and the electrostatic potential difference, respectively. Apart from external bias voltages that can be applied to the tip and sample, the potential difference term originates from the tip-sample work function difference, and/or from the existence of. electric charges and/or dipoles in the system under consideration. The basic principle of KPFM (Figure 1) consists in minimizing the electrostatic potential difference by providing a proper dc compensation bias ($V_{KPFM}$ or $V_{dc}$, applied in our case to the tip). It is common to assert that ($V_{tip}=V_{dc}$) matches the surface electrostatic potential ($V_s$). This simplification does not change in any way the validity of our arguments.

Probing the electrostatic interaction by KPFM relies on a lock-in detection scheme, in which a modulated bias Vmod of amplitude Vac and angular frequency $\omega_{mod}$ is added to the static dc voltage (Equ. S1 in the supplementary information). The total electrostatic force becomes the sum of three components, including a static term, and modulated components at the excitation frequency $\omega_{mod}$ and at $2\omega_{mod}$ referred herein to as first harmonic and second harmonic components:

$$F_{dc} = \frac{1}{2}C'_z\left[(V_{dc}-V_s)^2 + \frac{V_{ac}^2}{2}\right] (2a)$$

$$F_{\omega_{mod}} = C'_z(V_{dc}-V_s)V_{ac}\cos(\omega_{mod}t) (2b)$$

$$F_{2\omega_{mod}} = \frac{1}{4}C'_z V_{ac}^2 \cos(2\omega_{mod}t) (2c)$$

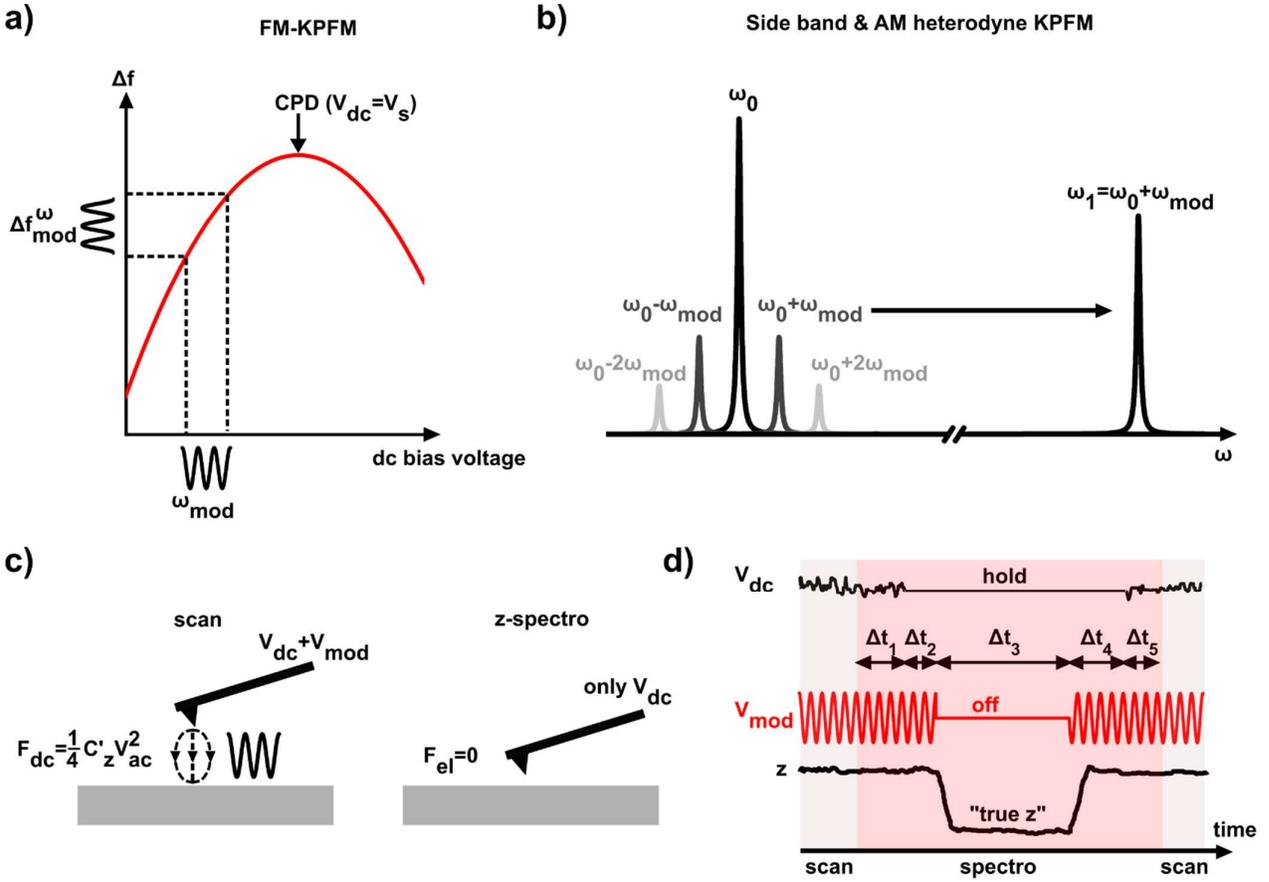

**Figure 1** KPFM modes. **a)** The cantilever frequency shift displays a quadratic dependence as a function of the tip-sample electrostatic potential difference. Applying an ac voltage (with angular frequency $\omega_{mod}$) causes periodic oscillations in the cantilever resonance frequency. In FM-KPFM, the surface potential ($V_s$) or contact potential difference (CPD) is obtained by minimizing (with a proper dc bias, $V_{dc}$) the amplitude of the modulated frequency shift at the first harmonic ($\Delta f\omega_{mod}$, main text, Equ. 3b). The second harmonic channel ($2\omega_{mod}$) yields a measurement of the tip-sample capacitance gradient z-derivative. **b)** Side-bands exist due to the frequency mixing between the electrical bias modulation and the cantilever mechanical oscillation (depicted here at its first eigenmode, $\omega_0$). $V_S$ is obtained by minimizing the amplitude of the first side bands ($\omega_0 \pm \omega_{mod}$). In AM-heterodyne KPFM, $\omega_{mod}$ is selected in order to shift the first right side-band ($\omega_{0+}\omega_{mod}$) to the second eigenmode frequency ($\omega_1$). **c)** In all cases, a static electrostatic force component proportional to the capacitance gradient and the square of the modulated bias amplitude remains. To perform a correct topographic measurement, one needs to apply the KPFM compensation dc bias ($V_{dc}$) without adding the modulated bias. **d)** Principle of the two-dimensional z-spectroscopy. In each pixel of the surface, the scan is stopped. The KPFM dc bias loop and the modulated bias are maintained during a first integration/stabilization delay ($\Delta t_1$). The dc bias is subsequently "frozen" by a sample and hold circuit. After a second delay ($\Delta t_2$) the bias modulation is switched off, during a time-lapse $\Delta t_3$. All components of the electrostatic force are now suppressed. $\Delta t_3$ is set to exceed the z-feedback time constant: the tip-sample distance has enough time to stabilize itself to its minimum. The bias modulation is turned on again, and the KPFM dc bias loop is reactivated after a fourth delay ($\Delta t_4$). The scan resumes after a last delay ($\Delta t_5$).

Instead of using the force, it is preferable to perform a force gradient detection (which minimizes the contribution of long-range electrostatic interaction) by using the cantilever frequency shift ($\Delta f$). The components of interest become:

$$\Delta f_{dc} = \frac{1}{2}C''_z\left[(V_{dc}-V_s)^2 + \frac{V_{ac}^2}{2}\right] (3a)$$

$$\Delta f_{\omega_{mod}} = C''_z(V_{dc}-V_s)V_{ac}\cos(\omega_{mod}t) (3b)$$

$$\Delta f_{2\omega_{mod}} = \frac{1}{4}C''_z V_{ac}^2 \cos(2\omega_{mod}t) (3c)$$

where C"z is the capacitance second z-derivative. In frequency-modulated KPFM (FM-KPFM), In frequency-modulated KPFM (FM-KPFM), the first harmonic component (Equ. 3b) is demodulated and injected as an input in the KPFM compensation potential feedback loop (Fig. 1a), which minimizes it by adjusting $V_{dc}$ to match $V_s$. Demodulating the second harmonic (Equ. 3c) yields a measurement of the tip-sample capacitance second z-derivative.

In side-band KFPM, the electrostatic information is obtained through the amplitude of lateral side bands (Fig. 1b). They stem from a heterodyning effect (or frequency mixing) between the electrical bias modulation and the cantilever mechanical oscillation (usually performed at the first eigenmode, with an angular frequency $\omega_0$). It can be simply shown (see the supporting information) that here, too, the signal is proportional to the force gradient. Amplitude modulated heterodyne-KPFM[11] (hereafter simply referred to as heterodyne-KPFM) is an interesting variant of side-band KPFM, in which the first side-band is shifted at the second cantilever eigenmode (frequency $\omega_1$) by performing the electrical excitation at $\omega_{mod}=\omega_1-\omega_0$. The sensitivity is thus boosted by performing the amplitude detection at the second resonance.

Whatever the detection scheme, the existence of a static electrostatic force component (Equ.2a) remains an issue. It is still with us, even when the tip dc bias matches finally the surface potential. This is a fundamental limitation of bias-modulated KPFM: to probe the surface potential one has no choice but to generate an additional "extrinsic" electrostatic force. This situation is illustrated in Fig. 1c. In other words, it is ultimately incorrect to claim that one nullifies the tip-sample electrostatic interaction by using an active KPFM loop.

As we have already pointed out, this should especially be a matter of concern if the tip-sample capacitance displays spatial variation; in that case, topographic artefacts seem unavoidable. Fortunately, these effects should be mitigated by working with small bias modulation amplitudes. One shall indeed remember here that the dc component of the force - or its gradient - is proportional to the square of the bias modulation amplitude. In the following, we will first check to what extent reducing the bias modulation amplitude allows performing a correct estimation of the thickness of TMD flakes grown by chemical vapor deposition (CVD) on silicon oxide. We will then demonstrate how "electrostatic free" measurements can be performed by implementing a dedicated z-spectroscopy protocol (Fig. 1c and 1d).

## 3. Results

With the idea of reducing as much as possible the modulated bias amplitude, heterodyne-KPFM shall be an asset. We are indeed going to see that, thanks to its sensitivity, it allows reducing $V_{ac}$ down to a few tens of mV while preserving a good potential resolution. It is also worth mentioning that, contrary to conventional amplitude-modulated KPFM, heterodyne KPFM is not affected by artefacts due to long-range electrostatic forces[11,12]. This stems from the fact that the electrostatic force component at the origin of the lateral side bands is proportional to the second capacitance z-derivative (see the supporting information). In short, heterodyne-KPFM somehow combines the advantages of AM-KPFM and FM-KPFM (in terms of sensitivity and lateral resolution, respectively). Nothing, however, is perfect: by shifting the first side band at the second eigenmode, we deprive ourselves of mapping the capacitive variations with a second harmonic demodulation.

Before going any further, it can thus be helpful to perform a first characterization of the samples under consideration by "conventional" FM-KPFM. In that case, we can rely on a dual harmonic demodulation to perform a simultaneous mapping of the surface potential and of the capacitance-related signal. In this work, the samples consist in $MoS_2$ flakes grown by CVD on doped-silicon substrates, covered by a 150nm thick thermal silicon oxide layer (Figure 2a). The CVD process parameters have been set to obtain MoS$_2$ monolayers. Basic topographic characterizations carried out by AFM in ambient conditions suggest that the TMD growth performed as expected. However, the effective stacking height deduced from these measurements (*ca.* 0.3 nm, see Figure S1 in the supplementary information) is much lower the expected value for one MoS$_2$ layer. This highlights the limitations of routine AFM topographic characterizations, when investigating atomically flat 2D materials on SiO$_2$.

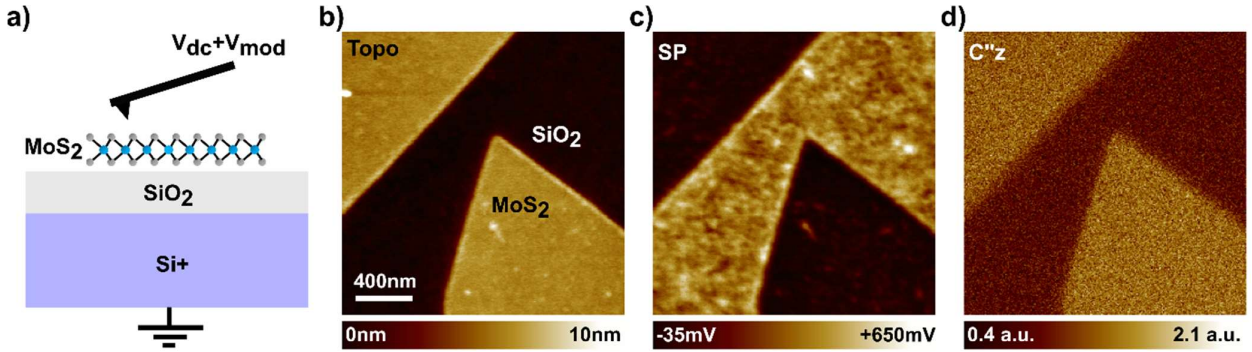

**Figure 2 a)** Scheme of the samples (CVD-grown MoS$_2$ flakes on SiO$_2$/Si) and experimental configuration. Both dc ($V_{dc}$) and modulated bias ($V_{mod}$) voltages are applied to the tip, the sample is grounded. KPFM data are presented as the tip dc bias (compensation potential, $V_{tip}=V_{dc}=V_{KPFM}$), that matches the surface potential (SP). **b,c,d)** Topography, surface potential and capacitance second z-derivative images acquired by FM-KPFM. 2000×2000nm, 300×300pixels. $V_{ac}$=620mV. $\omega_{ac}$=1140Hz.

Now, let us look at the flake thickness from the vantage of nc-AFM/FM-KPFM imaging. The images in Fig. 2b and Fig. 2c display the topography and the surface potential of a representative area of the sample, where MoS$_2$ flakes can easily be identified. The flakes display typical triangular shapes, and appear as dark patches (*i.e.* more negative) in the surface potential images. Interpreting that electrostatic contrast in terms of effective work function may not be straightforward, because the MoS$_2$ domains have been grown on an insulator. It is enough to state that they appear as negatively charged.

As highlighted before, it is just as important, to put it mildly, to investigate the capacitance variations when moving the AFM tip from the SiO$_2$ to the MoS$_2$. The second harmonic channel image (Fig. 2d) confirms indeed that this parameter displays a significant change: it is much larger over the TMD flake. Consequently, the AFM tip undergoes a stronger attractive force, that the z-feedback loop compensates by retracting the tip away from the surface. Indeed, an effective stacking height of approximately 6nm nanometers is deduced from the topographic image z-values histogram (not shown): this exceeds almost by one order of magnitude the value expected for one monolayer.

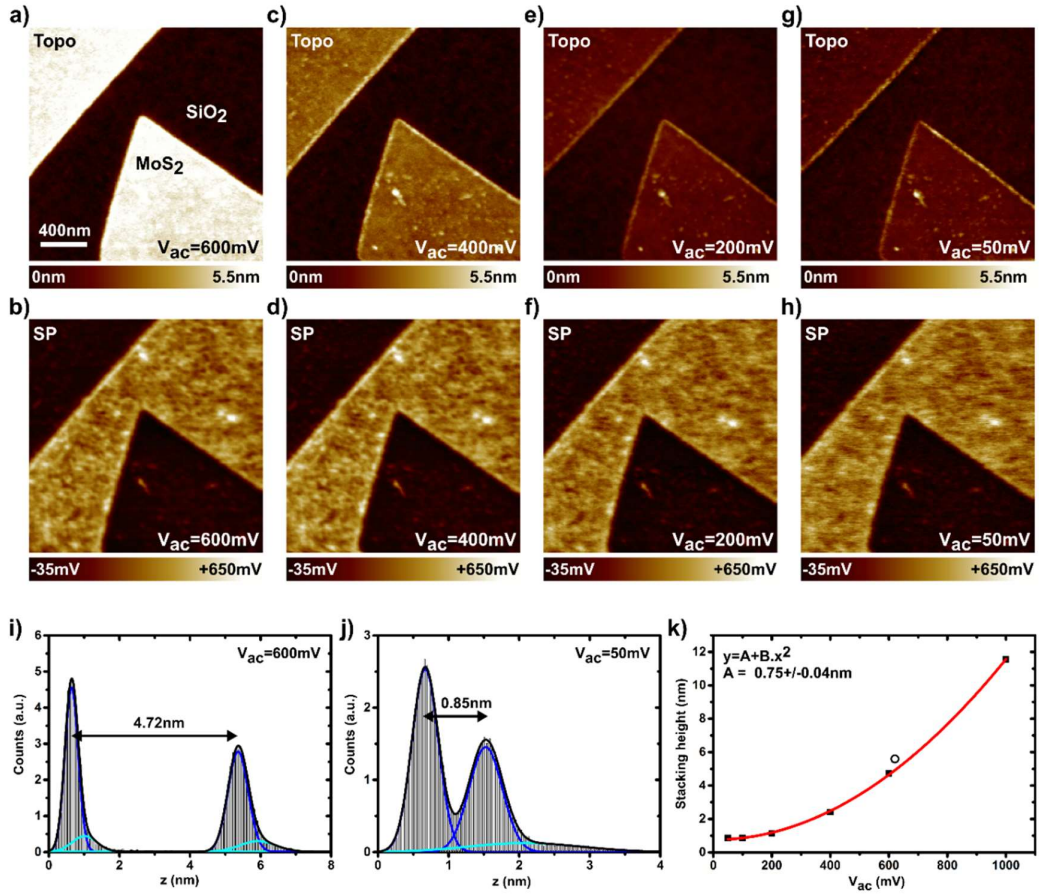

**Figure 3** Series of images acquired in heterodyne-KPFM (2000×2000nm, 300×300pixels), for decreasing bias voltage modulation amplitudes. $\omega_{ac}=\omega_1-\omega_0=394900Hz$. **a,c,e,g)** Topography **b,d,f,h)** Surface potential. **a,b)** $V_{ac}$=600mV. **c,d)** $V_{ac}$=400mV. **e,f)** $V_{ac}$=200mV. **g,h)** $V_{ac}$=50mV. **i,j)** Histograms of the images z-values for $V_{ac}$=600mV. (i) and $V_{ac}$=50mV (j). **k)** Black squares: MoS$_2$ stacking height on SiO$_2$ (deduced from z-histograms) as a function of $V_{ac}$. The open circle corresponds to the data obtained by FM-KPFM (shown in Fig. 2b). The red line shows the output of an adjustment of the data obtained by heterodyne-KPFM by a second order power law.

It is therefore absolutely imperative to reduce, as much as possible, the amplitude of the modulated bias. Now is the moment to take advantage of the heterodyne-KPFM capabilities. A series of images acquired with that mode on the same area than the FM-KPFM data is presented in Figure 3. During these experiments, the modulation bias amplitude has been progressively reduced (image per image) from a few hundreds of mV to a few tens of mV (four of six sets of data are displayed as images). The same colour scale has been used to map both topographic and potentiometric data, whatever the $V_{ac}$ value. It appears clearly that the apparent stacking height experiences a dramatic decrease when the modulation bias amplitude is reduced. In turn, the surface potential images display almost identical features (both in terms of potential levels and contrasts). This confirms – if it was needed – that the topographic dependence as a function of $V_{ac}$ is not related to an artefact in the surface potential compensation by the KPFM loop. Tip height histograms have been analysed for each image (two of them are shown in Fig. 3i and 3j), allowing to plot the dependence of the apparent stacking height as a function of $V_{ac}$. The data can be nicely adjusted using a quadratic equation, yielding a zero intercept value of 0.75±0.04nm. Doing so, one extrapolates the stacking height that would be measured in the absence of any perturbative electrostatic force. The interpolated value is fully consistent with what is expected for a single layer of $MoS_2$ in van der Waals interaction with the underlying substrate.

It is also worth mentioning that, even without performing a comprehensive analysis of the z-histogram $V_{ac}$-dependence, stacking height values below 1nm would have been deduced from the data acquired with modulation amplitudes equal or below 100mV (see Figure 3k). Thus, in our case, working with low $V_{ac}$ would have been sufficient to confirm the monolayer nature of the $MoS_2$ flake. However, there is no assurance that fixing the modulation amplitude below 100mV will always allow assessing the number of monolayers of any kind of TMD, stacked on any kind of oxide. Depending on the system under investigation, several parameters can indeed affect the tip-sample capacitance and how it varies when moving from the bare oxide to the TMD. To name just a few examples, there is every reason to believe that depending on the tip apex geometry, on the oxide thickness, and on the metallic, semiconducting or insulating nature of the TMD, the situation may differ.

In view of this, we need to go a step further. To do so, the only way is to remove all components of the electrostatic force during the topographic measurement (Fig. 1c and Fig. 1d ). At first glance, this seems impossible: the tip-sample potential difference cannot be compensated if one does not use a modulated bias to measure it. It is however possible to overcome this apparent deadlock by performing the surface potential measurement and electrostatic force cancellation in a sequential manner. In a first step, the KPFM controller is operated in a standard manner. Once the compensation potential has been generated by the KPFM feedback loop ($V_{dc}=V_{KPFM}$), it is maintained to a fixed value by using a sample and hold stage. Then, it is possible to switch off the modulated bias (because the output of the hold circuit stays equal to the compensation bias). At this stage, since $V_{dc}=V_s$ and $V_{ac}=0$, the electrostatic force has been totally cancelled. In practice, spectroscopic curves of the tip vertical displacement as a function of time are recorded in each image pixel. The spectroscopic acquisition is synchronized with two pulse trains generated by an arbitrary waveform generator. The first pulse channel drives the sample and hold stage (built around an analogic LF198 circuit from Texas Inst.). The second channel controls the application of the modulated bias via an analogic multiplication stage (based on a AD835 4-quadrant multiplier from Analog. Devices). Additional information can be found in the supplementary information (Fig. S2). Finally, it must be ensured that enough time is given to the topographic feedback loop to track properly the z-change after $V_{mod}$ switching off. The time-lapse during which $V_{ac}=0$ ($\Delta t_3$ in Fig.1d) is thus set to exceed the z-feedback time constant. Additional delays can be set at the beginning and the end of the spectroscopic sequence, in particular the first delay ($\Delta t_1$ in Fig.1d) can be extended to improve the signal-to-noise ratio on the compensation KPFM potential ($V_{dc}$).

The performances of this electrostatic compensated z-spectroscopy have been put to the test on a second $MoS_2$ sample; here again the CVD-process was performed to obtain monolayers on $SiO_2$. As we shall see, however, this sample feature defects that can be used to illustrate furthermore the crucial need to perform a fully "electrostatic-compensated" topographic imaging.

As before, it is preferable to carry out a preliminary scan in standard nc-AFM/FM-KPFM, the results of which are presented in Figure 4. Here again, the surface potential is shifted towards more negative values and the capacitance second derivative signal is in overall higher over the $MoS_2$ domains. So, it is not surprising that once again, the TMD appear much thicker in the z-images than it should be. Despite these similarities, the topographic images show significant differences with respect to the ones recorded on the former sample. The domain tip and some areas at the periphery along the edges display indeed a lower effective height, which is correlated with a specific contrast in the capacitance signal. More precisely, these areas are characterized by lower $C''_z$ values, which account for the topographic contrast.

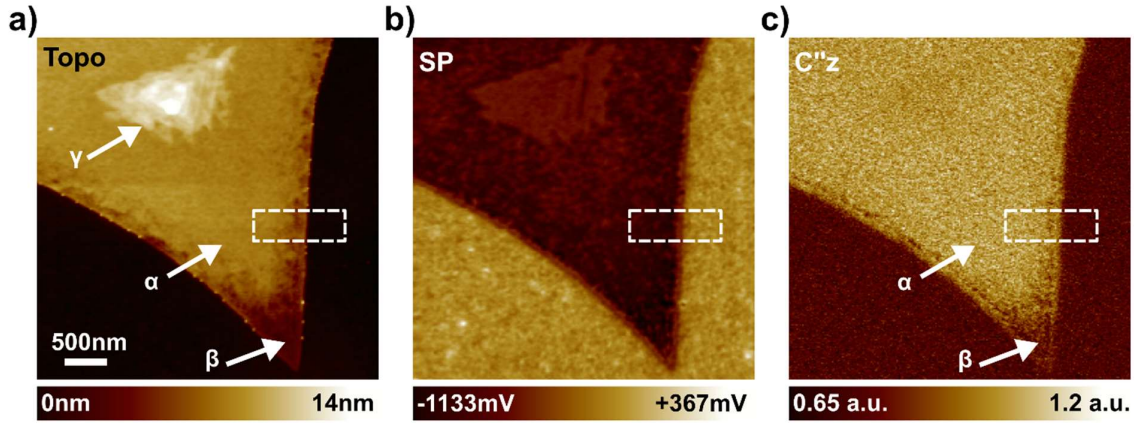

**Figure 4 a,bc)** Topography (a), surface potential (b) and capacitance second z-derivative (c) images acquired by FM-KPFM on a second $MoS_2$-$SiO_2$/Si sample. 4000×4000nm, 300×300pixels. $V_{ac}$=1.1V. $\omega_{ac}$=1140Hz. The arrows labeled by greek characters highlight three different areas within the TMD domain. α: apparent topographic level and capacitive signal (with respect to the bare substrate) similar to the ones of the former sample. β: "anomalous" area. γ: multilayer area. The dotted rectangle indicates the location of the area where a subsequent 2D z-spectroscopy will be performed (Fig.5).

If not careful, one might be tempted in view of the topography to conclude that the MoS$_2$ flake is covered in its most part by a contamination layer. We will see, thanks to the outputs of electrostatic-free imaging that the opposite happens.

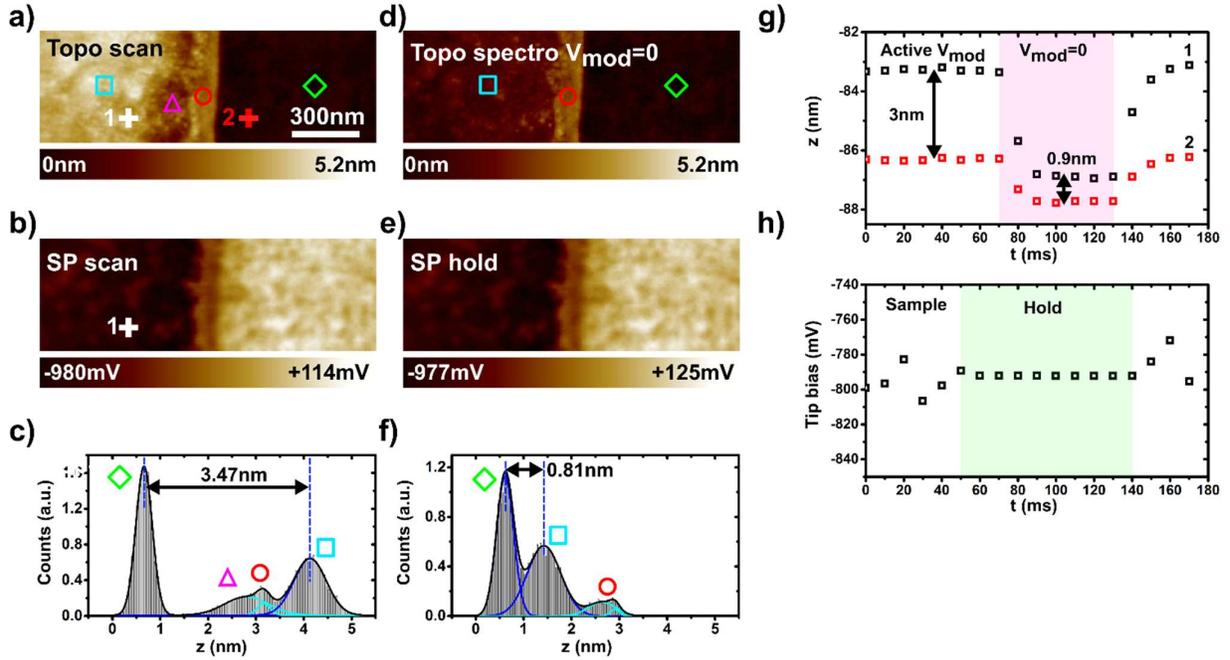

**Figure 5** 2D z-spectroscopy on the second MoS$_2$-SiO$_2$/Si sample. The data have been acquired on the location delimited by dotted-contours in Fig.4. **a,b)** Topographic (a) and potentiometric (b) images acquired during the scan with an active bias modulation (heterodyne KPFM, V$_{ac}$=500mV). 300×100pixels. 1500×500nm. **c)** Histogram of the topographic image z-values (V$_{ac}$=500mV). Square, triangle, circle and diamond symbols indicate the correspondence between the Gaussian peaks and different sample areas. **d,e)** Topographic (d) and potentiometric (e) images reconstructed from the 2D matrix of spectroscopic data for V$_{mod}$=0. **f)** Histogram of the topographic image z-values (V$_{mod}$=0). **g,h)** Spectroscopic curves of the tip height (g) and tip dc bias voltage (h). The data have been recorded at the locations highlighted by crosses labelled 1 (over the MoS$_2$ area) and 2 (over the SiO$_2$ substrate) in image (a). In (h), the dc bias curve is only shown for marker 1.

Figure 5 displays the outputs of a first 2D z- spectroscopy scan, which has been performed in heterodyne- KPFM on rectangular area that spreads on both sides of the TMD edge (highlighted by dotted contours in Fig.4). The first series of images (Fig. 5a,b) correspond to the data acquired in the scan phase during which both the modulated bias (V$_{mod}$) and the compensation bias (V$_{dc}$) are applied (see Fig 1.d). The second series of images (Fig 5d,e) has been obtained from the matrix of spectroscopic curves (a selection of curves is shown in Fig. 5g,h ), by mapping the z values (Fig.5d) and tip compensation bias (Fig.5e) values recorded during the time-lapse where V$_{mod}$=0, and after allowing time to the AFM tip to reach a minimum z-level. During each spectroscopic acquisition, this time-interval falls between t=100ms and t=120ms, as shown in Fig. 5g. Two curves acquired at different locations (over the TMD and over the bare substrate) are presented (Fig .5g), along with a curve of the surface potential value (recorded over the TMD). It is easy to see that the relative z-level difference between both curves strongly decreases when the modulation bias is switched off. Accordingly, the topographic image contrast undergoes a dramatic change (compare Fig. 5a and Fig. 5d), while the surface potential channel (Fig. 5b vs Fig. 5e) stays almost perfectly constant; that is, the tip dc bias is fixed to the proper value by the sample and hold circuit when $V_{mod}=0$. By carrying out an analysis based on the histograms of the z-levels, it turns out that the average stacking height of the TMD on the underlying substrate varies from ca. 3.5nm when $V_{ac}=500mV$ to ca 0.8nm when $V_{ac}=0mV$. For the sake of completeness, we also repeated a series of measurements following our first protocol, *i.e.* by recording a series of data for decreasing $V_{ac}$ values (see Figure S3 in the supplementary information). The outputs of both experiments are fully consistent, but it is worth noting that in this case, reducing the bias down to 100mV is not sufficient to reach a sub-nm stacking height value (see Fig. S3). As previously evoked, it turns out that the best measurements are achieved when the electrostatic force is fully suppressed.

The sub-nm stacking height value deduced from the data acquired under zero bias modulation confirms that, here too, the CVD process yielded in average a $MoS_2$ monolayer. However, we still have to understand what the origin of the odd topographic contrasts is (*i.e.* the contrasts observed under the influence of the dc electrostatic component). A careful examination of the data might provide a clue to understanding the nature of the phenomenon at play. As highlighted before, the $MoS_2$ flake displays specific features near its edges. In Figure 5a, the area highlighted by a triangle lies lower than the domain interior (labeled by a square), and the average z-level just at the edge (indicated by a circle) falls in between. Remarkably, the first two areas become completely levelled when one performs the correct topographic height measurement (see Fig. 5d, and compare z-histograms In Fig. 5c and Fig. 5f ), and only a narrow band at the edge continues to appear as a locally raised elevation.

As the last scan was performed on a restricted sample area, one might wonder if our last observation applies to the entire flake. A subsequent 2D z-spectroscopic mapping has therefore been performed on a larger scale (Figure 6). The conclusion is clear: the topography is almost completely perfectly levelled when the electrostatic contributions are removed from the force field. This unambiguously demonstrates that the whole flake is constituted by a single $MoS_2$ monolayer, except for a central multilayer area (labeled $\gamma$ in Figure 4), and for the protrusion at the flake edge (labeled by a red circle in Figure 5d, and highlighted by an arrow in Figure 6b). A question then remains: what is the phenomenon at the origin of the contrasts observed on this sample in "standard" nc-AFM/KPFM imaging?

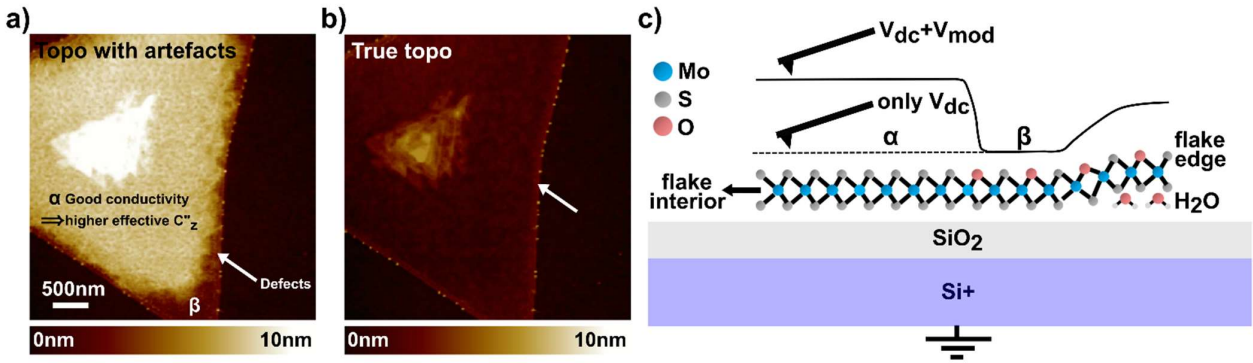

**Figure 6** "Electrostatic free" topographic imaging of the second a second $MoS_2$-$SiO_2$/Si sample. 2D z-spectroscopy. Heterodyne KPFM , $V_{ac}$=700mV. 4000×4000nm, 200×200pixels. **a,b)** Images of the z-levels reconstructed from the 2D matrix of spectroscopic data. **a)** z-levels recorded with an active modulation bias. The defect-free area labelled α feature an apparent topographic level (with respect to the bare substrate) similar to the one of the former sample. The area labeled β is probably partially oxidized, which results in a smaller capacitance second derivative (see Figure 4c), and consequently in an apparent lower topographic level. **b)** z-levels recorded when $V_{mod}$ is switched off. The arrow pinpoints the existence of a protrusion at the edge, tentatively attributed to intercalated water molecules or other molecular contaminants. **c)** Scheme illustrating our hypothesis. The oxygen atoms in the oxidized β area have been here arbitrarily positioned in substitutional positioning (replacing top sulfur atoms), but the real situation may differ from this artwork[27].

Some defects must be at play. CVD is well-established as a powerful technique to grow high-quality TMD monolayers, in particular $MoS_2$[13-16]. However, it relies on a large number of experimental parameters (growth temperature, annealing and cooling rates, inert gas flux and pressure, type and amount of sulfur and metal precursors, substrate preparation, use or not of molecular growth promoters[17], furnace geometry, etc.). For each parameter, small perturbations can lead to drastic modifications in the properties of the grown TMD flakes such as shape, size, crystalline quality, doping and number of layers. Growth inhomogeneity not only concerns different batches, but most importantly, at the single synthesis level, significant flake-to-flake and intra-flake inhomogeneity are common. Typical defects include atomic scale defects (in particular sulfur vacancies) and grain boundaries[14,15], charge fluctuations associated with the former defects and with trapped charges in the substrate, symmetry breaking at edges[18], oxidation sites[19], adsorbates, growth nucleation sites and/or multilayer areas, molecules intercalated between the TMD and substrate and mechanical strain[20]. Studies using Raman and photoluminescence mapping[15,16,16,21-26] often report that the edges and tips of crystalline TMD flakes behave differently from the core region. In particular, in CVD-based samples, it is well-established that mechanical strain resulting from the difference in thermal coefficient of the TMD and the substrate is a major cause of intra-flake inhomogeneity of the optical properties[21,24-26]. This strain, accumulated during the cooling stage of the CVD synthesis, is differentially relaxed at edges, tips and grain boundaries. In addition, it is also well recognized that at the post-growth stage, water intercalates between TMDs and hydrophilic substrates[22] (such as the $SiO_2$ used in this study).

In view of what precedes, interesting conclusions can be drawn. At the flake interior, existing defects (such as sulfur vacancies) have a negligible effect on the real sample corrugation. Otherwise, the stacking height measured with the electrostatic-free mode would not match so closely the expected value for a monolayer. Only the flake edges display noticeable topographic sur-elevations (sample area highlighted by red circles in Figures 5 and 5d). Since the samples have been handled in air prior to in-vacuum measurements, these last features (protrusions at the edges) may possibly be attributed to intercalated water at the $SiO_2/MoS_2$ interface. It seems much less likely that hydration of the oxide substrate extends deep into the flake's interior: again, the existence of a water layer would go against the measured "electrostatic free" stacking height.

As a matter of fact, the most plausible hypothesis, is that the flake periphery is partially oxidized. Consequently, it displays a lower conductivity and acts less efficiently as top "metallic" electrode that enhances the effective tip-sample capacitance gradient. Therefore, the areas of lower altitude observed in the "standard" topographic/KPFM images (labelled β in Figures 4 and 6) correspond to oxidized parts of the $MoS_2$ flake. Conversely, the non-defective area at the interior (labelled α) displays a higher conductivity, and a higher apparent stacking height under the application of the modulated bias. This very counter-intuitive situation is highlighted in Figure 6.

This scenario is fully consistent with the fact that the contaminant intercalation is known to progress from the flake edges and inward. In other words, the contaminated area should be located at the flake periphery rather than at its interior. Yet to be definitely confirmed, the hypothesis of a partial oxidation is also very likely, for the configuration of many kinds of oxidation sites is compatible with the absence of noticeable impact on the topography (off course we discuss here the "true" topography, *i.e.* the one that is probed in the "electrostatic-free" mode). For instance, it has been shown[27] that oxygen can exist as substitutional impurities on the $MoS_2$ surface (as tentatively depicted in Figure 6c).

Our interpretation is also strengthened by the fact that almost identical stacking height are deduced from measurements performed in the same conditions on the interior of the second sample flake, and on the first (apparently) defect-free sample. For instance, measurements performed in "standard" heterodyne-KPFM with a modulated bias amplitude of 500mV yield stacking values of ca. 3.6nm for the first sample (see the data in Figure 3k), and ca. 3.5nm for the interior of the second sample flake (Figure 5c).

Overall, investigating this inhomogeneous sample allowed us to exemplify how our approach not only solves the issue of accurate thickness measurement but can also help studying inhomogeneity in MoS$_2$ flakes from a new perspective.

## Conclusion

We have introduced a simple approach for artefact-free topographic measurements by nc-AFM/KPFM, based on the acquisition of spectroscopic curves of the tip-surface distance in close-loop configuration. Its implementation requires only using a few basic analog circuits to synchronously maintain the compensation bias and switch off the modulation voltage; it should also be easy to develop numeric counterparts directly integrated in last generation digital scanning probe microscope controllers. Like in the case of graphene[10], we have shown that conventional nc-AFM/KPFM measurements cannot assess correctly the stacking height of 2D transition metal dichalcogenides deposited on silicon-oxide substrates. This further highlights just how the variations of the tip-sample capacitance can affect the topographic measurement, despite the CPD compensation by the KPFM controller. Decreasing the modulated bias amplitude reduces the error, in that sense heterodyne-KPFM appears as a promising alternative to FM-KPFM, thanks to its higher sensitivity. However, as demonstrated, error-free measurements can only be performed if all components of the electrostatic force are removed. This warning is to be taken seriously by the community of researchers working with non-contact AFM under UHV. Indeed, the high quality factors under vacuum boost the cantilever sensitivity to electrostatic forces, and the electrostatic-induced topographic artefacts should accordingly be exacerbated. Our results also prompt us to be very careful when applying nc-AFM/KPFM to defect identification at the TMD/oxide interface, since we have shown that the topographic contrast can counterintuitively be affected by the variations of the capacitive gradient. By cancelling the total electrostatic force, one can perform a z-regulation based on tip-sample interactions stemming only from van der Waals forces (as long as the nc-AFM is operated in a true non-contact regime). Beyond the case of single TMD monolayers, this "van der Waals force microscopy" should become a tool of choice for the characterization of 2D van der Waals heterostructures, for their opto-electronic properties depend critically upon the number of layers stacked at the atomic scale, and the quality of the interfaces that they form with technological oxide substrates.

## Experimental

Noncontact-AFM (nc-AFM) experiments were performed with a ScientaOmicron VT-AFM setup in ultra-high vacuum at room temperature. The scanning probe microscope is driven by a Matrix control unit (ScientaOmicron). Topographic imaging was realized in FM mode (FM-AFM) with negative frequency shifts of a few Hz and vibrational amplitudes of a few tens of nm. A dual channel lock in (7280, Signal Recovery) was used for FM-KPFM, while heterodyne-KPFM measurements were carried out with a numeric lock-in (MFLI, Zurich Instruments). Pt/Ir-coated silicon cantilevers (EFM, Nanosensors, resonance frequency in the 45–115 kHz range) were annealed in situ to remove atmospheric contaminants. The KPFM compensation voltage $V_{dc}$ was applied to the cantilever (tip bias $V_{tip} = V_{dc}$). The KPFM data are presented as $V_{dc}$ images also referred to as KPFM potential or surface potential images for simplicity.

$MoS_2$ domains were synthesized by chemical vapor deposition (CVD) as in Ref. 28, based on a process initially adapted from[14]. Briefly, sulfur and $MoO_3$ powders were annealed in a nitrogen flow within the quartz tube of a tubular furnace using their respective positions to adjust their temperature (in the 180-220°C and 700-750°C ranges for S and $MoO_3$ respectively). The Si/SiO2 substrate, covered with a thin film of perylene-3,4,9,10-tetracarboxylic acid tetrapotassium salt (PTAS) acting as seed promoter was placed at the same temperature as the $MoO_3$ source and kept at the maximal temperature for 10 minutes.

## Authors contributions

B.G. implemented the z-spectroscopic mode, carried out the nc-AFM/KPFM experiments with A.A., and analysed the data. The $MoS_2$ samples were CVD-grown by N.U. under the supervision of V.D. B.G. wrote the manuscript, with inputs from V.D. and N.U. for the CVD growth description and discussion of the defects.

## Acknowledgments

Financial support by the Agence Nationale de la Recherche (France) with the Matra2D project (ANR-20-CE24-0017) is gratefully acknowledged. CEA-Licsen thanks Quentin Cogoni for his contribution to the $MoS_2$ synthesis during his internship.

**Side bands in KPFM**

The attractive electrostatic force between the tip and the surface is:

$$F = \frac{1}{2} C'_z [(V_{dc} - V_s) + V_{mod}]^2 \text{ with } V_{mod} = V_{ac} \cos(\omega_{mod} t) \quad (S1)$$

$V_{dc}$, $V_{mod}$ and $V_s$ stand for the dc compensation bias, the modulated bias (angular frequency $\omega_{mod}$, amplitude $V_{ac}$) and the surface potential, respectively. The electrostatic force is the sum of three components:

$$F_{dc} = \frac{1}{2} C'_z \left[ (V_{dc} - V_s)^2 + \frac{V_{ac}^2}{2} \right] \quad (S2a)$$

$$F_{\omega_{mod}} = C'_z (V_{dc} - V_s) V_{ac} \cos(\omega_{mod} t) \quad (S2b)$$

$$F_{2\omega_{mod}} = \frac{1}{4} C'_z V_{ac}^2 \cos(2\omega_{mod} t) \quad (S2c)$$

To understand the nature of the side bands, one need only to consider the fact that the capacitance gradient oscillates with the same angular frequency that the AFM cantilever ($\omega_0$ for the first eigenmode). Actually, since the cantilever oscillates in a nonharmonic potential, higher harmonics are needed to describe its motion. A Fourier series can therefore be used to describe the capacitance gradient.

$$C'_z = \frac{k_0}{2} + \sum_{n=1}^{+\infty} k_n \cos(n\omega_0 t) \quad (S3)$$

Consequently, multiple spectral components appear in the electrostatic force. In particular, if we restrict ourselves to the first harmonic (n=1), two components emerge that are:

$$F_{\omega_0 \pm \omega_{mod}} = \frac{1}{2} k_1 V_{ac} (V_{dc} - V_s) \cdot [\cos(\omega_0 - \omega_{mod}) + \cos(\omega_0 + \omega_{mod})] \quad (S4a)$$

$$F_{\omega_0 \pm 2\omega_{mod}} = \frac{1}{8} k_1 V_{ac}^2 \cdot [\cos(\omega_0 - 2\omega_{mod}) + \cos(\omega_0 + 2\omega_{mod})] \quad (S4b)$$

Therefore, electrostatic signals can be detected by demodulating the cantilever amplitude at the side bands $\omega_0 \pm \omega_{mod}$ and $\omega_0 \pm 2\omega_{mod}$. Last, by expanding the capacitance gradient in a Taylor series, it can be shown that the Fourier coefficient amplitude at n=1 is proportional to the first z-derivative of the capacitance gradient (see the work by Axt and co-workers [S1] for a detailed calculation):

$$k_1 \equiv C''_z \quad (S5)$$

**Reference**

[S1] Axt A., Hermes I. M., Bergmann V. W., Tausendpfund N. and Weber S. A. L. Know your full potential: Quantitative Kelvin probe force microscopy on nanoscale electrical devices *Beilstein J. Nanotechnol.* **9** 1809 (2018).

**First MoS$_2$/SiO$_2$-Si sample: complementary characterizations by AM-AFM in ambient conditions**

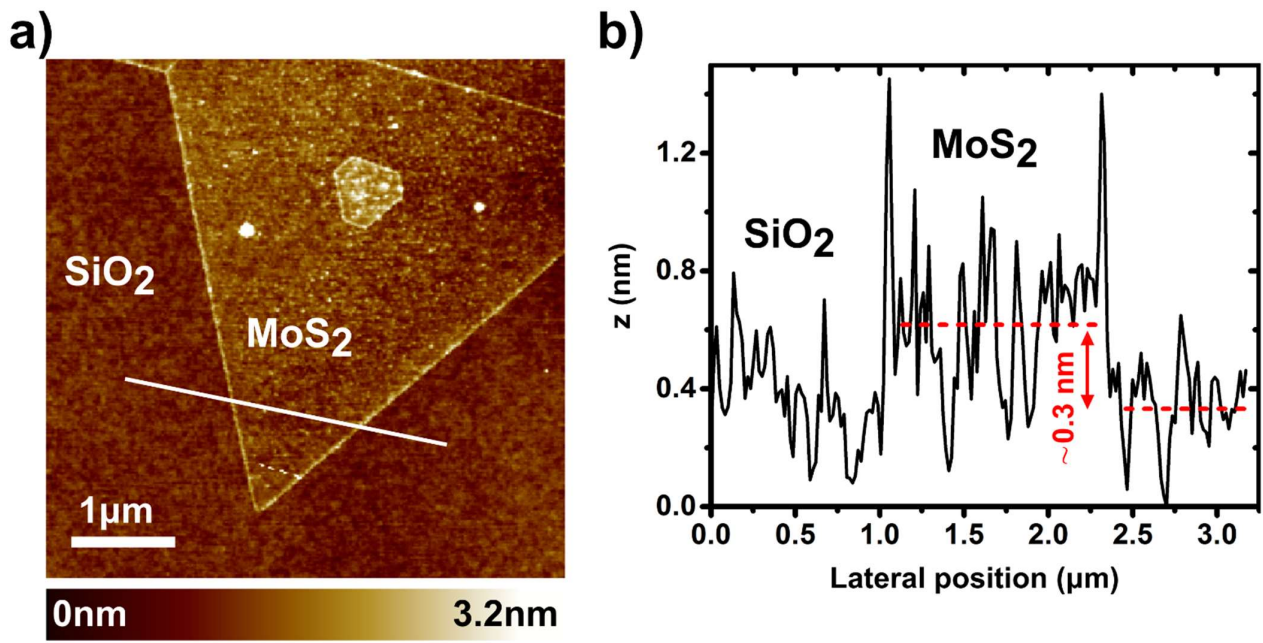

**Figure S1** a) Large scale (5×5μm, 304×304 pixels) AFM image recorded in ambient conditions (AFM imaging in amplitude modulation, with an Icon setup from Bruker) on the first MoS$_2$-SiO$_2$/Si sample. b) Cross section topographic profile corresponding to the path highlighted by a white line in the topographic image. The apparent stacking height (*ca.* 0.3nm) that would be deduced from that profile falls well below 0.7nm. The mean z-levels over the MoS$_2$ flake and over the SiO$_2$ substrate (red dotted lines) have been determined by averaging the z-levels for 1.1μm<x< 2.8μm and 2.4μm<x<,3.2μm respectively.

**z-spectroscopy implementation: experimental scheme**

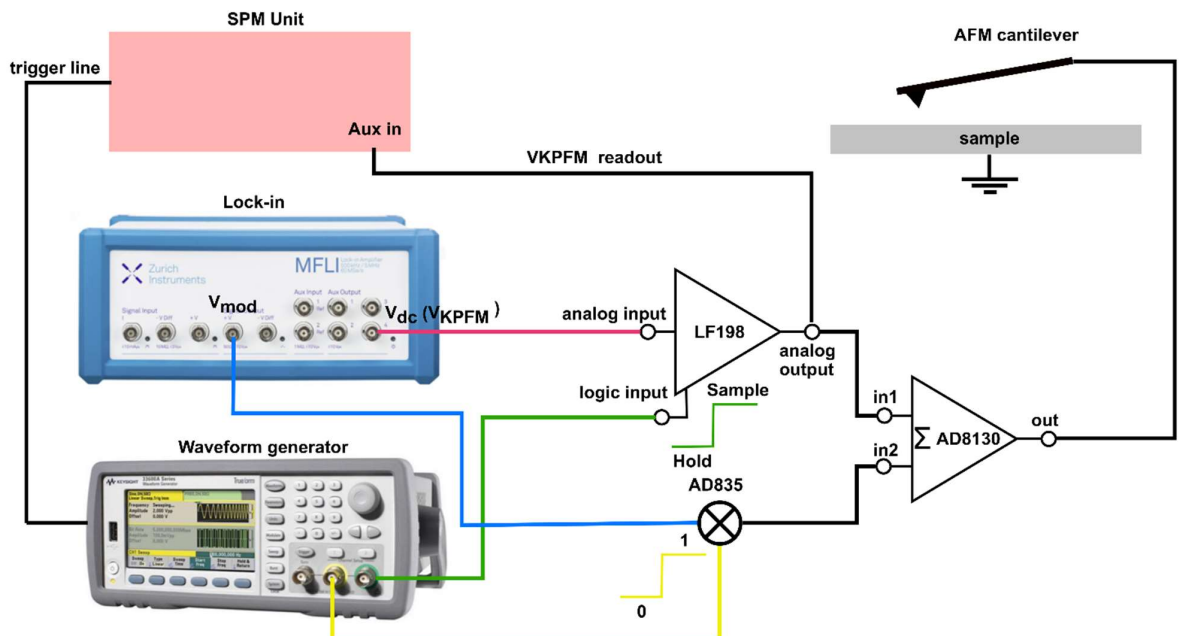

**Figure S2** Scheme of the z-spectroscopy implementation. The dc compensation potential generated by the KPFM controller (MFLI unit from Zurich Instruments) is fed to the input of a sample and hold circuit (based on an analogic LF198 component from Texas Inst). In the "hold state", this circuit maintains $V_{dc}$ at a constant value (average of the last values integrated over a selected time-constant). The modulated bias ($V_{mod}$) is multiplied by 1 (on state) or 0 (off state) by using a 4-quadrant analog multiplier circuit (AD835 from Analog. Devices). The static and modulated bias are summed with an AD8130 stage (from Analog. Devices), and applied to the AFM cantilever. The pulse trains used to drive the sample and hold circuit (logic input) and the multiplier are generated by a dual channel arbitrary waveform generator (Keysight 38622A). The spectroscopic acquisition (z-curves as a function of time, see the main text) is synchronized with the arbitrary waveform generatot channels by means of a trigger line.

**Second MoS$_2$/SiO$_2$-Si sample: topography and surface potential as a function of V$_{ac}$**

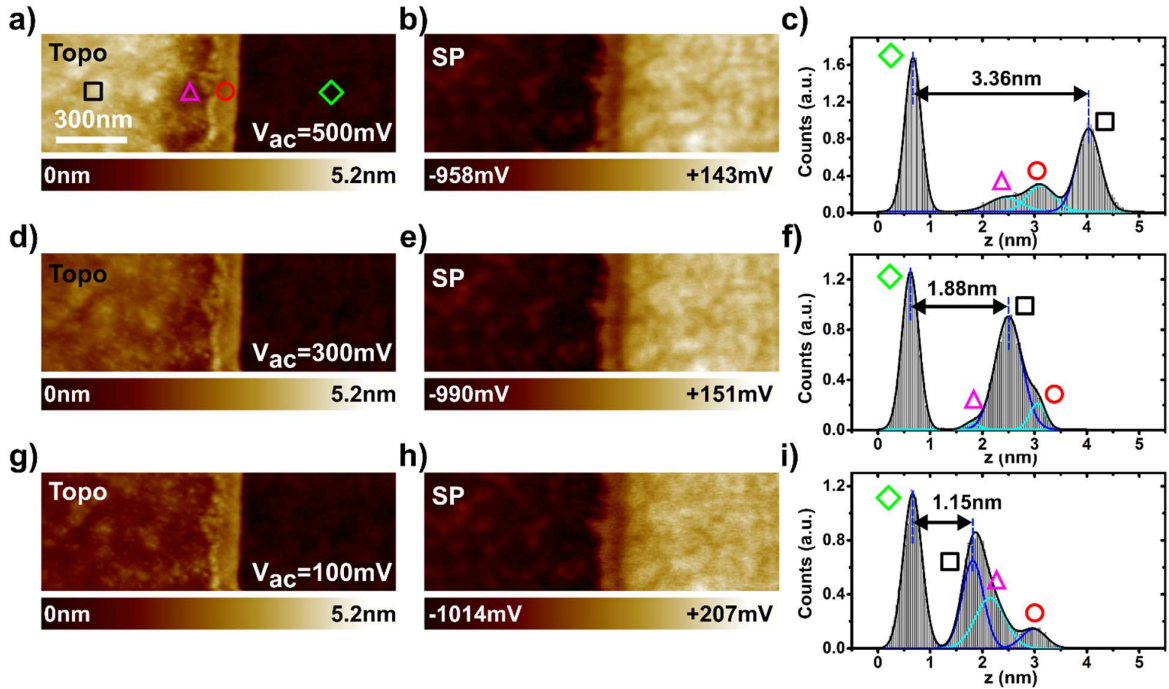

**Figure S3** Series of images acquired on the second MoS$_2$-SiO$_2$/Si sample in heterodyne-KPFM (1500×500nm, 300×100pixels), for decreasing bias voltage modulation amplitudes. **a,d,g)** Topography **b,e,h)** Surface potential. **a,b)** V$_{ac}$=500mV. **d,e)** V$_{ac}$=300mV. **g,h)** V$_{ac}$=100mV. **c,f,i)** Histograms of the images z-values for V$_{ac}$=500mV (c), 300mV (f) and 100mV (i).